\documentclass[preprint, showpacs,preprintnumbers,amsmath,amssymb,nofootinbib]{revtex4}
\usepackage{mathrsfs}
\usepackage{amsmath}
\usepackage[colorlinks, linkcolor=blue, citecolor=red, urlcolor=red]{hyperref}
\usepackage{amsfonts}
\usepackage{latexsym}
\usepackage{amsfonts}
\usepackage{graphicx}
\usepackage{epsf}
\usepackage{dcolumn}
\usepackage{bm}

\newcommand{\bea}[1]{\begin{eqnarray}\label{#1}}
\newcommand{\eea}{\end{eqnarray}}

\def\gsim{ \lower .75ex \hbox{$\sim$} \llap{\raise .27ex \hbox{$>$}} }
\def\lsim{ \lower .75ex \hbox{$\sim$} \llap{\raise .27ex \hbox{$<$}} }

\begin{document}
 \title{Re-examining the role of curvature in the slowing down acceleration scenario}

\author{Jianmang Lin, Puxun Wu and Hongwei Yu }
\address
{Center of Nonlinear Science and Department of Physics, Ningbo
University,  Ningbo, Zhejiang, 315211 China }

\begin{abstract}
By incorporating the curvature $\Omega_k$ as a free parameter, it has been found that  the tension between the high redshift cosmic microwave background (CMB) shift parameter $R(z^{\ast})$ data and the low redshift Type Ia supernova (SNIa) and baryonic acoustic oscillation (BAO) data from the combination of Sloan Digital Sky Surver (SDSS) and Two-Degree Field Galaxy Redshift Survey (2dFGRS) can be  ameliorated,  and  both SNIa+BAO and SNIa+BAO+CMB favor that the decelerating parameter $q(z)$  shows a rapid variation in the sign at the small redshift. In this paper, with the Monte Carlo Markov chain method,  we re-examine the evolutionary behavior of $q(z)$ using  the latest observational data including the Union2 SNIa,  BAO, and   CMB data ($R(z^{\ast})$, $l_{A}(z^{\ast})$, $z^{\ast}$) from Wilkinson Microwave Anisotropy Probe 7-yr (WMAP7). For the BAO data, four different data sets  obtained from  the  Two-Degree Field Galaxy Survey,
the combination of SDSS and 2dFGRS, the WiggleZ Dark Energy
Survey and the  Baryonic Oscillation Spectroscopic Survey, are used.  Except for the spatially flat case constrained by SNIa+ the WiggleZ BAO data, both SNIa and other BAO+SNIa favor that the present cosmic acceleration is slowing down irrespective of whether the spatial curvature is included or not. However, once  the WMAP7 CMB data is included,  observations favor strongly the Lambda cold dark matter model, a spatially flat universe, and a speeding-up of the cosmic acceleration. Therefore, the inclusion of spatial curvature seems to has no effect on alleviating the tension between SNIa+BAO and CMB in clear contrast to the previous work in the literature.

\end{abstract}

\maketitle

\section{Introduction}\label{sec1}
Two independent Type Ia supernova (SNIa) observation groups discovered firstly in 1998 that the Universe is undergoing an accelerating expansion~\cite{Perlmutter1999, Riess1998}. This discovery was  further confirmed by observations on the cosmic large scale structure~\cite{Eisenstein, Tegmark} and the cosmic microwave background radiation~\cite{Spergel}. At the same time, the observations also indicated that the cosmic  phase transition from deceleration to acceleration occurs only in the near past.
Interestingly,  using the Constitution~\cite{Hicken} and Union2~\cite{Amanullah} SNIa along with the baryonic acoustic oscillation (BAO) distance ratio between redshifts $z=0.2$ and $0.35$~\cite{Percival2010} and  the Chevalier-Polarski-Linder(CPL) parametrization~\cite{Chevallier} for the equation of state (EOS) of dark energy, the authors in Refs.~\cite{Shafieloo, Li2011} recently found  that the acceleration of the cosmic expansion  might be slowing down. However, once the  CMB data is included, their results turn out to be supporting that the universe is undergoing an accelerating expansion with an increasing acceleration.   Thus, there appears some tension between the low redshift data (SNIa+BAO) and the high redshift (CMB) one. This tension can be reduced by various methods, which yield different results on the cosmic expanding history~\cite{Li2011}. Through combining different  SNIa and BAO data, Gong et al~\cite{Gong} found that the systematics in the data sets does influence the fitting results and leads to different evolutional behavior of the decelerating parameter.  The effect of different parametrizations of the EOS of dark energy  has also been  studied~\cite{Shafieloo, Gong, Li2010}. In addition, it has been found that  the  accelerating cosmic expansion might be transient by performing a cosmographic evaluation~\cite{Guimaraes} or using a bin method for $q(z)$~\cite{Cai}.

More recently,   using the SNIa (Constitution and Union2~\cite{Hicken, Amanullah}), BAO  distance ratio from the joint analysis of the Two-Degree Field Galaxy Redshift Survey (2dFGRS) and Sloan Digital Sky Survey (SDSS) data~\cite{Percival2010}  and CMB shift parameter from Wilkinson Microwave Anisotropy (WMAP)~\cite{Hinshaw, Komatsu2011}, Cardenas and Rivera~\cite{Cardenas} found that the tension  between the low and high redshift data can be ameliorated effectively if the curvature $\Omega_k$ is incorporated as a free parameter,  and both SNIa+BAO and SNIa+BAO+CMB favor  a rapid variation in sign of $q(z)$ between $z \sim 0.5$ and $ 0$.

In this paper, we plan to reanalyze the evolutionary behavior of $q(z)$ with  the curvature $\Omega_k$  included as a free parameter in our discussion. The latest SNIa, BAO and CMB data will be used. Different from~\cite{Cardenas} where only BAO distance ratio is considered, we use four different BAO data sets to probe the systematics. These include the data at the low-redshift  $z=0.106$~\citep{Beutler} from 6-degree Field Galaxy Survey (6dFGS),  at redshifts z = 0.35 and 0.20 from the SDSS and the 2dFGRS~\cite{Percival2010},   at redshifts z = 0.44, 0.6 and 0.73 from the baryon acoustic peak released by the WiggleZ Dark Energy Survey~\citep{Blake},   and  at $z=0.57$ from the Baryonic Oscillation Spectroscopic Survey (BOSS) ~\citep{Sanchez}.  Furthermore, for CMB data, we use the measurements of  three derived quantities from WMAP7~\cite{Komatsu2011}: the shift parameter $R(z^*)$ and the acoustic index $l_a(z^*)$ at the recombination redshift $z^*$, rather than only the shift parameter, since these three quantities can give unbiased information on dark energy parameters relative to the full CMB analysis as demonstrated in~\cite{HongLi}.

\section{OBSERVATIONAL DATA }

For SNIa data, we use the Union2 data set released by the Supernova
Cosmology Project (SCP) Collaboration~\cite{Amanullah},  which consists of 557 data points.
To constrain a theoretical model from  SNIa, we minimize the $\chi^2$ value of distance
 modulus
\begin{eqnarray}\label{Eq1}
\chi^2=\sum^{557}_{i=1}\frac{[\mu(z_{i})-\mu_{obs}(z_{i})]^{2}}{\sigma^{2}_{\mu i}},
\end{eqnarray}
where $\mu(z)\equiv5\log_{10}[d_{L}(z)/Mpc]+25$
is the theoretical value of distance modulus, and $\mu_{obs}$ is the
corresponding observed one.  $d_{L}$ is the luminosity distance and it is  defined as
\begin{eqnarray}
d_{L}(z)=\frac{1+z}{\sqrt{|\Omega_k|}}\frac{c}{H_{0}}S_k\bigg[\sqrt{|\Omega_k|}\int^{z}_{0}\frac{dz'}{E(z')}\bigg],
\end{eqnarray}
where the dimensionless Hubble parameter $E(z)=H(z)/H_{0}$, and $S_k(x)$ is defined as $x$, $\sin(x)$ or $\sinh(x)$ for $k = 0$, $+1$, or $-1$, respectively. For a dark energy model with the equation of state $w(z)$, one has
\begin{eqnarray}
E^{2}(z)=\Omega_{m}(1+z)^{3}+\Omega_{k}(1+z)^{2}+\Omega_{de}\exp\bigg(3\int^{z}_{0}\frac{1+w(z')}{1+z'}dz'\bigg),
\end{eqnarray}
where $\Omega_m$ and $\Omega_{de}$ are the dimensionless density parameters of matter and dark energy, and $\Omega_{de}=1-\Omega_{m}-\Omega_{k}$. Thus, for the CPL parametrization, $\textit{w}=\textit{w}_{0}+\textit{w}_{1}z/(1+z)$, with $w_0$ and $w_1$ being two model parameters,
we get
\begin{eqnarray}
E^{2}(z)=\Omega_{m}(1+z)^{3}+\Omega_{k}(1+z)^{2}+\Omega_{de}(1+z)^{3(1+w_{0}+w_{1})}\exp\bigg(-\frac{3w_{1}z}{1+z}\bigg).
\end{eqnarray}
Since $H_0$ is a nuisance parameter, we marginalize over it with a flat prior and then obtain
\begin{eqnarray}
\chi^2_{SNIa}(\bold{p})=\sum_{i=1}^{557}\frac{\alpha_i^2}{\sigma_i^2}-\frac{(\sum_i \alpha_i/\sigma_i^2-\ln 10/5)^2}{\sum_i1/\sigma_i^2}-2\ln \bigg(\frac{\ln10}{5}\sqrt{\frac{2\pi}{\sum_i1/\sigma_i^2}}\bigg),
\end{eqnarray}
where $\alpha_i=\mu_{obs}(z_i)-25-5\log_{10}[H_0d_L(z_i)]$, and $\bold{p}$ denotes the fitting parameters in the model.

For  BAO data, we use the measurements from the 6dFGS (hereafter
 BAO1) \cite{Beutler}, the combination of  SDSS and 2dFGRS~\cite{Percival2010} (hereafter BAO2),  the WiggleZ dark energy survey (hereafter BAO3)
  \cite{Blake} and the BOSS (hereafter BAO4) \cite{Sanchez}.
For BAO1,  the
distance ratio $d_{z}=r_{s}(z_{d})/D_{V}(z)$ is measured at redshift $z=0.106$ and the result is $d^{obs}_{0.106}=0.336\pm0.015$~\cite{Beutler},
where the  effective distance is
\begin{eqnarray}
D_{V}(z)=\bigg[\frac{d^{2}_{L}(z)}{(1+z)^{2}}\frac{z}{H(z)}\bigg]^{1/3},
\end{eqnarray}
the drag redshift $z_d$ is fitted as~\cite{Eisenstein1998}
\begin{eqnarray}
z_d=\frac{1291(\Omega_mh^2)^{0.251}}{1+0.659 (\Omega_mh^2)^{0.828}}[1+b_1(\Omega_bh^2)^{b_2}]
\end{eqnarray}
with $h=H_0/100$, $b_1=0.313(\Omega_mh^2)^{-0.419}[1+0.607(\Omega_mh^2)^{0.674}]$ and $b_2=0.238(\Omega_mh^2)^{0.223}$,
and the comoving sound horizon is
\begin{eqnarray}
r_{s}(z)=\int^{\infty}_{z}\frac{c_{s}(x)dx}{E(x)},
\end{eqnarray}
where the sound speed is
$c_{s}(z)=1/\sqrt{3[1+\overline{R}_{b}/(1+z)]}$ with $\overline{R}_{b}=3\Omega_{b}/(4\times2.469\times10^{-5})$,
and $\Omega_{b}$ being the dimensionless baryon matter energy density. The constraint from BAO1 can be obtained by minimizing
\begin{eqnarray}
\chi^2_{B1}=\frac{(d_{0.106}-0.336)^2}{0.015^{2}}
\end{eqnarray}

From SDSS and   2dFGRS, Percival et al.~\cite{Percival2010} measured two distance ratios at redshifts $z=0.2$
and $z=0.35$ and obtained $d^{obs}_{0.2}=0.1905\pm0.0061$, $d^{obs}_{0.35}=0.1097\pm0.0036$. Using
\begin{eqnarray}
\chi^2_{B2}=\sum^{2}_{i,j=1}\triangle d_{i}\;C^{-1}_{dz}(d_{i},d_{j})\;\triangle d_{j},
\end{eqnarray}
one can obtain the result from BAO2. Here $\triangle d_{i}=d_i-d_i^{obs}$ and $C_{dz}$ is the covariance matrix for two parameters $d_{0.2}$ and $d_{0.35}$ given in \cite{Percival2010}.

 BAO3, containing three data points at $z=0.44, 0.6$ and $0.73$,  is given by
the WiggleZ dark energy survey~\cite{Blake}. The acoustic parameter rather than the distance ratio
\begin{eqnarray}
A(z)=\frac{D_{V}(z)\sqrt{\Omega_{m}H^{2}_{0}}}{cz}
\end{eqnarray}
is measured  in~\cite{Blake}. The constraint on model parameters can be obtained by using
\begin{eqnarray}
\chi^2_{B3}=\sum^{3}_{i,j=1}\triangle A_{i}C^{-1}_{A}(A_{i},A_{j})\triangle A_{j}\;,
\end{eqnarray}
where the covariance matrix $C_{A}(A_{i},A_{j})$
is given in Table 2  in~\cite{Blake}.

Recently, BOSS~\cite{Sanchez}  released a BAO data $A(0.57)=0.444\pm0.014$, which can be used to test theoretical model by minimizing
\begin{eqnarray}
\chi^2_{B4}=\frac{(A(0.57)-0.444)^2}{0.014^2}\;.
\end{eqnarray}

Using $\chi^2_{BAO}=\chi^2_{B1}+\chi^2_{B2}+\chi^2_{B3}+\chi^2_{B4}$, we can obtain the result from the combination of all BAO data sets.

Except for the low redshift SNIa and BAO data, we also use the high redshift CMB information by implementing the Wilkinson microwave anisotropy probe 7 year (WMAP7) data. Since, when the full WMAP7 data are applied, some more parameters depending on inflationary models need to be added, the ability to constrain dark energy models will be limited. Thus,
we use three derived quantities~\cite{Komatsu2011}: the shift parameter $R(z^{\ast})$ and
the acoustic index $l_{A}(z^{\ast})$ at the recombination redshift $z^{\ast}$.
\begin{eqnarray}
R(z^{\ast})=\frac{\sqrt{\Omega_m}}{\sqrt{|\Omega_k|}}S_k\bigg(\sqrt{|\Omega_k|}\int^{z^{\ast}}_0\frac{dz}{E(z)}\bigg)=1.725\pm0.018,
\end{eqnarray}
\begin{eqnarray}
l_{A}(z^{\ast})=\frac{\pi d_{L}(z^{\ast})}{(1+z^{\ast})r_{s}(z^{\ast})}=302.09\pm0.76,
\end{eqnarray}
where the redshift $z^{\ast}$ is given in ~\cite{Hu1996}
\begin{eqnarray}
z^{\ast}=1048[1+0.00124(\Omega_{b})^{-0.738}][1+g_{1}(\Omega_{m}h^{2})^{g_{2}}]=1090.04\pm0.93
\end{eqnarray}
with
\begin{eqnarray}
g_{1}=\frac{0.0783(\Omega_{b}h^{2})^{-0.238}}{1+39.5(\Omega_{b}h^{2})^{0.763}},
\end{eqnarray}
\begin{eqnarray}
g_{2}=\frac{0.560}{1+21.1(\Omega_{b}h^{2})^{1.81}}.
\end{eqnarray}
Thus, the $\chi^{2}_{CMB}$ can be expressed as
\begin{eqnarray}
\chi^{2}_{CMB}(\textbf{p})=\Sigma^{3}_{i,j=1}\triangle x_{i}C^{-1}_{CMB}(x_{i},x_{j})\triangle x_{j},
\end{eqnarray}
where the three parameters $x_{i}=[R(z^{\ast}),l_{A}(z^{\ast}),z^{\ast}]$,
$\triangle x_{i}=x_{i}-x^{obs}_{i}$ and covariance matrix $C_{CMB}(x_{i},x_{j})$
is taken from Table 10 in~\cite{Komatsu2011}. It has been found that these CMB quantities  can give similar constraints on dark energy parameters compared with the full CMB power spectrum~\cite{HongLi}.

\section{results}
We use the Monte-Carlo Markov Chain (MCMC) method to explore the model parameter
space \textbf{p}. The MCMC method randomly chooses values of
parameters $\textbf{p}$, calculate $\chi^{2}$ and determines
whether to accept or reject the set of parameters $\textbf{p}$
using the Metropolis-Hastings algorithm method~\cite{Lewis2002}.

To probe the properties of dark energy and the cosmic expansion  history,
we analyze the constraints on the CPL parametrization and then reconstruct the evolutionary behavior of the decelerating parameter
$q(z)=-\ddot{a}a/\dot{a}^{2}$.  The $q(z)$ gives information on the expansion speed and acceleration  since its sign tells us whether  the expansion is accelerating or decelerating, and its first derivative shows whether the acceleration is slowing down or speeding up.

\subsection{spatially flat case}
\begin{figure}[h!]
\centering
\includegraphics[width=1\linewidth]{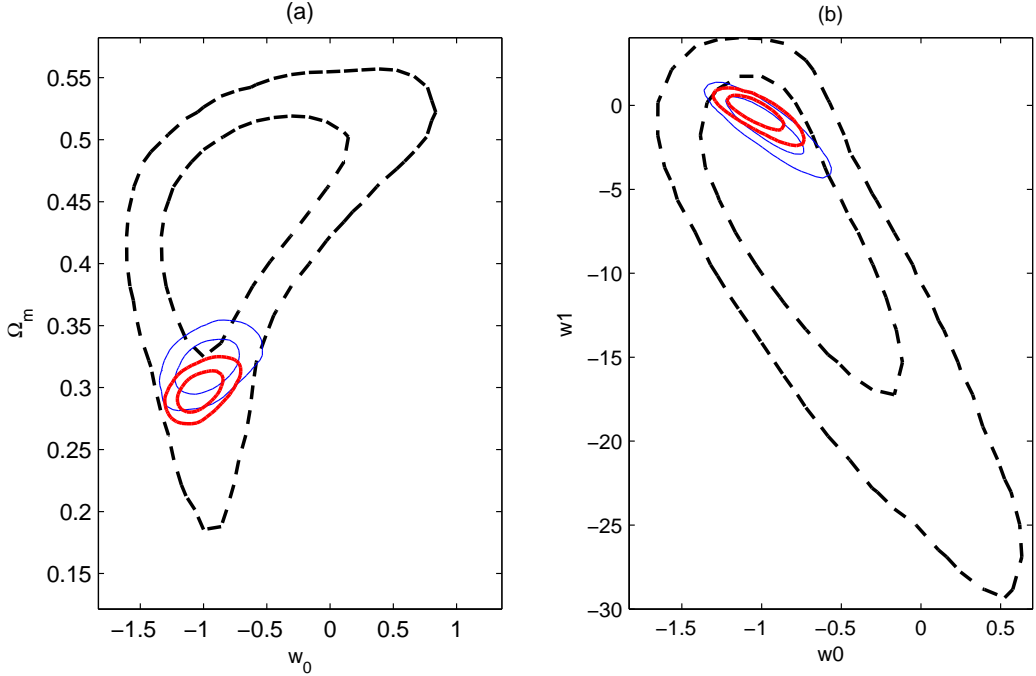}
\caption{The $1\sigma$ and $2\sigma$ contours in $w_{0}-\Omega_{m}$~~(a)
and $w_{0}-w_{1}$~~(b) planes for the  flat CPL model.  The dashed, solid and thicken solid lines shows the results from SNIa, SNIa+BAO and SNIa+BAO+CMB, respectively. }
\label{Fig1}
\end{figure}

\begin{figure}[h!]
\centering
\includegraphics[width=1\linewidth]{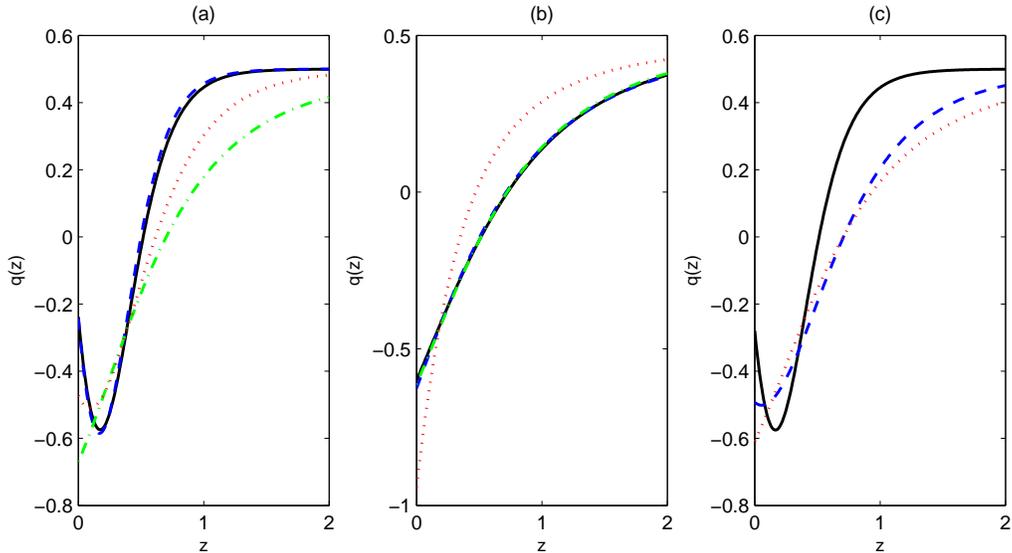}
\caption{The best fit evolutionary curves of  the deceleration parameter  for the flat CPL model. In (a), the solid, dashed,   dot-dashed and dotted lines
represent the results obtained from SNIa+BAO1, SNIa+BAO2, SNIa+BAO3 and SNIa+BAO4, respectively. In (b),  the solid, dashed,   dot-dashed and dotted lines
represent the results obtained from SNIa+BAO1+CMB, SNIa+BAO2+CMB, SNIa+BAO3+CMB and SNIa+BAO4+CMB, respectively. In (c),
the solid, dashed and dotted lines show the results from   SNIa, SNIa+BAO   and SNIa+BAO+CMB, respectively.
}
\label{Fig2}
\end{figure}

The constraints on model parameters $\Omega_{m}$, $w_0$ and $w_1$
are given in Tab.~(\ref{Tab1}) and Fig.~(\ref{Fig1}),  and the evolutionary behaviors of $q(z)$ are shown in Fig.~(\ref{Fig2}).
It is easy to see that  the Union2 SNIa alone only has a weak constraint on model parameters. The best fit line of $q(z)$ shows that a slowing down of expansion acceleration, whose phase transition occurs at $z\sim0.3$, is favored by SNIa.  Adding BAO
data into the analysis, one can find that BAO1 and BAO2 have an almost negligible effect on results. However, when BAO4 is included,  the model parameters, especially $w_1$, are tightened  apparently,  but a slowing down of the expansion acceleration is still preferred.  Once  BAO3 is considered, a tighter constraint is obtained and, different from the cases of BAO1, BAO2 and BAO4, BAO3+SNIa favors that the cosmic acceleration is  speeding up. Thus, different BAO data sets seem to give different results, which implies that there exists a tension between  them.  Combining all BAO data and SNIa data (SNIa+BAO1+BAO2+BAO3+BAO4), we find that the phase transition from speeding up to slowing down remains to be favored, but the transition redshift is very close to zero.

When the WMAP7 CMB data  is included, the slowing down phenomenon  of the expansion acceleration, obtained in BAO1+SNIa, BAO2+SNIa, BAO4+SNIa and BAO+SNIa,  disappears in agreement with what was obtained in Ref~\cite{Li2011, Gong, Cardenas} where the CMB shift parameter and BAO distance ratio between  redshift $0.2$ and $0.35$  are used.  However, BAO3+SNIa and BAO3+CMB+SNIa give a very consistent result.   So, the tension between the low redshift BAO+SNIa data and the high redshift CMB data may not always exist. In addition, BAO4+CMB+SNIa seem to  show a slight difference on constraining model parameters and it allows  larger $\Omega_m$ and  $w_1$, and a more negative $w_0$.  This difference  can also be seen from the evolutionary curve of $q(z)$ in the middle panel of Fig.(\ref{Fig2}).

\subsection{spatially curved case}
\begin{figure}[h!]
\centering
\includegraphics[width=1\linewidth]{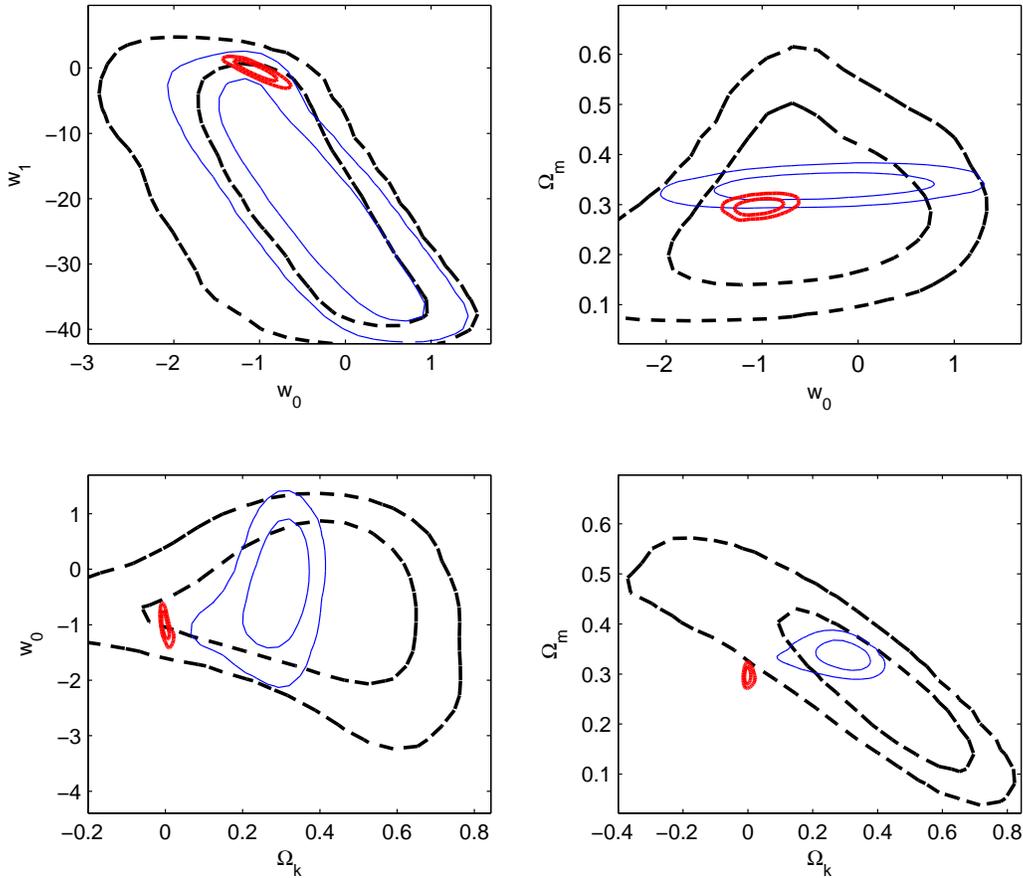}
\caption{The $1\sigma$ and $2\sigma$ contours  for the spatially curved  CPL model.  The dashed, solid and thicken solid lines shows the results from SNIa, SNIa+BAO and SNIa+BAO+CMB, respectively. }
\label{Fig3}
\end{figure}

\begin{figure}[h!]
\centering
\includegraphics[width=1\linewidth]{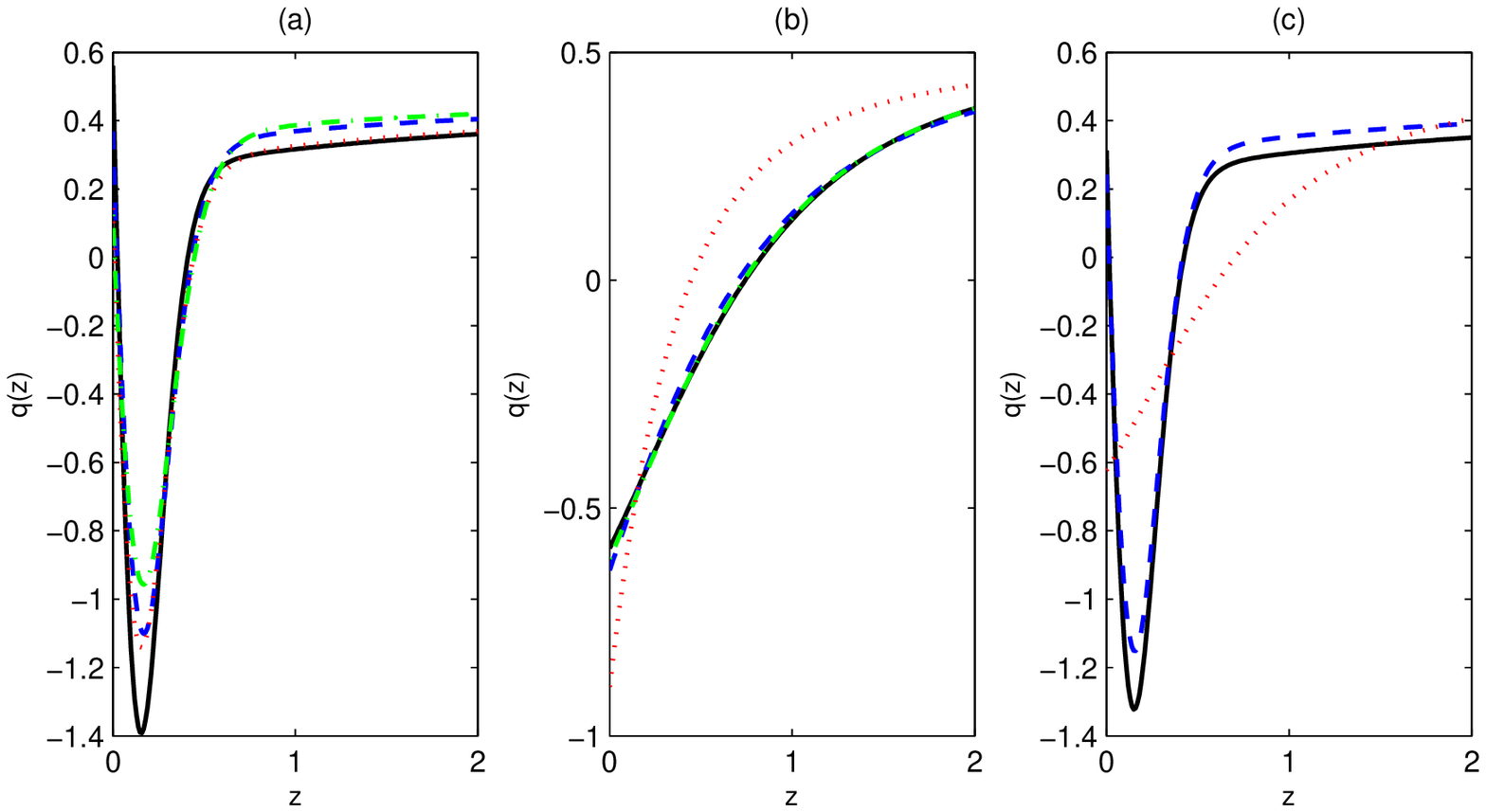}
\caption{The best fit evolutionary curves of  the deceleration parameter  for the CPL model with spacial curvature included. In (a), the solid, dashed,   dot-dashed and dotted lines
represent the results obtained from SNIa+BAO1, SNIa+BAO2, SNIa+BAO3 and SNIa+BAO4, respectively. In (b),  the solid, dashed,   dot-dashed and dotted lines
represent the results obtained from SNIa+BAO1+CMB, SNIa+BAO2+CMB, SNIa+BAO3+CMB and SNIa+BAO4+CMB, respectively. In (c),
the solid, dashed and dotted lines show the results from   SNIa, SNIa+BAO   and SNIa+BAO+CMB, respectively.}
\label{Fig4}
\end{figure}

In Ref.~\cite{Cardenas},  it has been found that the spatial curvature has a significant effect on reconstructing the evolution of $q(z)$ and alleviating the tension between low redshift data and high redshift one  since  both SNIa+CMB shift parameter+BAO distance ratio, and SNIa+BAO distance ratio   favor a slowing down acceleration when the spatial curvature is considered, which is just opposite to the flat case.  Thus, here, we also study the effect of  spatial curvature. The results are shown in Tab.~(\ref{Tab2}) and Fig.~(\ref{Fig3}). The corresponding $q(z)$ evolutionary curves are given in Fig.~(\ref{Fig4}).    We find that SNIa and SNIa+BAO (including SNIa+BAO1, SNIa+BAO2, SNIa+BAO3, SNIa+BAO4 and SNIa+BAO) give very consistent constraints on model parameters. This differs from the flat case where BAO3 leads to a different result.   An open universe is favored slightly although $\Omega_k=0$ is still allowed at the $1\sigma$ confidence level. Therefor, the tension between different BAO data sets obtained in flat case disappears when the spatial curvature is taken into account. The left and right  panels of Fig.~(\ref{Fig4}) show that the cosmic acceleration has entered a slowing down era and $q_0>0$ seems to be  favored, which means that the   accelerating cosmic expansion may be a transient phenomenon and the universe may have re-entered the decelerating phase.  In addition, we find that a very large $w_0$, $w_0\sim-0.5$, is given by SNIa and BAO data.

 When the WMAP7 CMB data is further added into our analysis,   the best fit results  change drastically. In this case, $\Omega_k$ is very close to zero and thus a flat universe is supported strongly.   Except for the SNIa+BAO4+CMB case, which favors  a  phantom-like dark energy at the present  with a crossing of $-1$ line in the near past, observations prefer a Lambda cold dark matter model since $w_0$ is very close to $-1$ and $w_1$ is very small.   The middle and right panels of Fig.~(\ref{Fig4}) show that a speeding up expansion is supported when  CMB is included, which is the same as the flat case, but is different from what was obtained  in Ref.~\cite{Cardenas} where only the CMB shift parameter is considered.  Thus, comparing the results in spatially flat and curved cases, we find that the inclusion spatial curvature leads to somewhat more serious tension between SNIa+BAO and CMB, although it alleviates it between different BAO data sets.

\begin{table}[!h]
\tabcolsep 0pt \caption{The marginalized $1\sigma$ constraints on flat CPL model by different observational data.} \vspace*{-12pt}
\begin{center}\label{Tab1}
\def\temptablewidth{1\textwidth}
{\rule{\temptablewidth}{1pt}}
\begin{tabular*}{\temptablewidth}{@{\extracolsep{\fill}}ccccc}
Data set &$\chi^{2}_{min}$  &$\Omega_{m}$  &$w_{0}$  &$w_{1}$
 \\   \hline  SN  &541.4544     &$0.4160^{+0.1054}_{-0.2908}$
 &$-0.8901^{+1.2419}_{-0.5818}$   &$-5.2281^{+6,9081}_{-19.0726}$   \\
 \hline  SN+BAO1  &541.4586 &$0.4162^{+0.1105}_{-0.2901}$
   &$-0.8430^{+1.3289}_{-0.6734}$   &$-5.4582^{+6.9106}_{-20.5981}$  \\
 SN+BAO2    &542.1974    &$0.4246^{+0.0925}_{-0.2780}$
 &$-0.8600^{+1.2845}_{-0.6166}$   &$-5.8181^{+7.1387}_{-18.5244}$  \\
SN+BAO3   &542.6020  &$0.2923^{+0.0550}_{-0.0465}$
&$-1.1019^{+0.4478}_{-0.2809}$   &$-0.2590^{+2.0659}_{-2.6811}$  \\
SN+BAO4   &541.9226 &$0.3580^{+0.0576}_{-0.0476}$
 &$-1.0102^{+0.5587}_{-0.4321}$   &$-1.9484^{+3.1705}_{-5.4024}$  \\
 SN+BAO    &549.3644    &$0.3168^{+0.0303}_{-0.0304}$  &$-0.9689^{+0.3580}_{-0.3172}$
   &$-1.1910^{+2.0452}_{-2.7636}$           \\
   \hline  SN+BAO1+CMB  &543.2798      &$0.2789^{+0.0303}_{-0.0276}$
   &$-1.0178^{+0.2359}_{-0.2243}$   &$0.0011^{+0.8396}_{-1.4796}$  \\
 SN+BAO2+CMB    &544.2428    &$0.2772^{+0.0315}_{-0.0253}$
 &$-1.0390^{+0.2433}_{-0.2081}$     &$0.1145^{+0.7107}_{-1.3773}$  \\
SN+BAO3+CMB   &543.1966     &$0.2808^{+0.0268}_{-0.0260}$
   &$-1.0323^{+0.2477}_{-0.2094}$   &$-0.0072^{+0.8337}_{-1.4231}$  \\
SN+BAO4+CMB   &548.4828    &$0.3201^{+0.0294}_{-0.0487}$
    &$-1.4142^{+0.6588}_{-0.1487}$   &$1.4904^{+0.1667}_{-3.7770}$ \\
    SN+BAO+CMB   &554.8478     &$0.2939^{+0.0269}_{-0.0231}$
&$-1.0464^{+0.2848}_{-0.3129}$   &$-0.2106^{+1.6254}_{-1.7849}$   \\

       \end{tabular*}
       {\rule{\temptablewidth}{1pt}}
       \end{center}
       \end{table}

\begin{table}[!h]
\tabcolsep 0pt \caption{The marginalized $1\sigma$ constraints on  CPL model with the spatial curvature considered.} \vspace*{-12pt}
\begin{center}\label{Tab2}
\def\temptablewidth{1\textwidth}
{\rule{\temptablewidth}{1pt}}
\begin{tabular*}{\temptablewidth}{@{\extracolsep{\fill}}cccccc}
Data set &$\chi^{2}_{min}$  &$\Omega_{m}$  &$w_{0}$  &$w_{1}$  &$\Omega_{k}$
 \\   \hline  SN  &541.0736      &$0.2880^{+0.3524}_{-0.2248}$
   &$-0.5205^{+2.4921}_{-2.3291}$   &$-18.8115^{+23.1852}_{-21.1880}$
     &$0.3689^{+0.4096}_{-1.0917}$  \\

      \hline  SN+BAO1  &541.1354      &$0.3022^{+0.2505}_{-0.1745}$
   &$-0.2766^{+1.6595}_{-2.6467}$   &$-21.4317^{+24.0151}_{-18.5661}$
   &$0.3503^{+0.3206}_{-0.6947}$  \\
 SN+BAO2    &542.0138  &$0.3486^{+0.1778}_{-0.2164}$
 &$-0.3875^{+1.8056}_{-2.2403}$   &$-16.4537^{+18.4183}_{-23.5444}$
 &$0.2467^{+0.4139}_{-0.3979}$  \\
SN+BAO3   &541.2002  &$0.3067^{+0.0598}_{-0.0495}$
 &$-0.6575^{2.1457}_{1.4439}$   &$-15.3458^{+16.9569}_{-24.6526}$
   &$0.3273^{+0.1185}_{-0.3430}$  \\
SN+BAO4   &541.2142 &$0.3628^{+0.0509}_{-0.0526}$
     &$-0.6103^{+1.9350}_{-1.2468}$  &$-13.2337^{+14.0533}_{-26.7524}$
       &$0.2084^{+0.1598}_{-0.4707}$  \\
       SN+BAO    &545.2002    &$0.3395^{+0.0452}_{-0.0438}$
  &$-0.5456^{+2.0612}_{-1.5226}$ &$-17.1917^{+18.2527}_{-22.8049}$
   &$0.2827^{+0.1075}_{-0.4803}$  \\
 \hline  SN+BAO1+CMB  &543.2394  &$0.2788^{+0.0369}_{-0.0284}$
 &$-1.0048^{+0.3765}_{-0.3936}$  &$-0.0947^{+1.5520}_{-2.6395}$
 &$-0.0011^{+0.0250}_{-0.0140}$  \\
 SN+BAO2+CMB    &544.2288  &$0.2776^{+0.0314}_{-0.0270}$
  &$-1.0490^{+0.3535}_{-0.3089}$     &$0.1714^{+1.2135}_{-3.0902}$
   &$0.0004^{+0.0244}_{-0.0159}$  \\
SN+BAO3+CMB   &543.1082  &$0.2777^{+0.0314}_{-0.0251}$
 &$-1.0248^{+0.4204}_{-0.3656}$   &$-0.0287^{+1.4668}_{-2.8580}$
 &$-0.0009^{+0.0263}_{-0.0147}$  \\
SN+BAO4+CMB   &548.1942    &$0.3374^{+0.0247}_{-0.0677}$
 &$-1.3977^{+0.7424}_{-0.1526}$   &$1.4873^{+0.1623}_{-4.6124}$
  &$-0.0031^{+0.0272}_{-0.0089}$  \\
SN+BAO+CMB   &554.8226  &$0.2935^{+0.0293}_{-0.0236}$
   &$-1.0583^{+0.4387}_{-0.3867}$ &$-0.2181^{+1.7094}_{-2.9104}$
     &$-0.0010^{+0.0250}_{-0.0124}$ \\
       \end{tabular*}
       {\rule{\temptablewidth}{1pt}}
       \end{center}
       \end{table}

\section{Conclusion}
In this paper, using the MCMC method,  we reconstruct the evolutionary behavior of the decelerating parameter $q(z)$ from the latest observational data including the Union2 SNIa, BAO, and  CMB data from WMAP7. For  BAO data, four different kinds of data obtained from the  6dFGS,
the combination of SDSS and 2dFGRS, the WiggleZ dark energy
survey and the BOSS are used. In our analysis, the CPL parametrization for the EOS of dark energy is considered. For a spatially  flat universe, we find that there is  tension between different BAO data sets since BAO1+SNIa,   BAO2+SNIa and  BAO4+SNIa support a slowing down of the acceleration of the cosmic expansion, while  BAO3+SNIa does not.   When the WMAP7 CMB data ($R(z^{\ast})$, $l_{A}(z^{\ast})$ and $z^{\ast}$)  is added into our discussion,  the slowing down phenomenon disappears and rather a speeding up is favored. Thus, there is also  tension between BAO+SNIa and CMB, except for the case of BAO3 since BAO3+SNIa and CMB+BAO3+SNIa give a very consistent constraint on model parameters.   By  incorporating the spatial curvature as a free parameter, we find that  SNIa and SNIa+BAO seem to support an open universe, the accelerating cosmic expansion  may be a transient phenomenon, and the tension between different BAO datasets is alleviated effectively.  However, when the CMB data from WMAP7 is included,  observations favor strongly a Lambda cold dark matter model and  a spatially flat universe.  Meanwhile, observations with CMB included prefer that the cosmic acceleration is speeding up, which is consistent with the flat case and what was obtained in Refs.~\cite{Shafieloo, Li2011}, but is different from what was obtained in \cite{Cardenas} where it was found that in a spatially curved universe, the SNIa+BAO distance ratio+CMB shift parameter gives a consistent result with SNIa+BAO distance ratio and both of them favor a slowing down of the cosmic acceleration. Comparing our  results obtained  in the flat and  curved cases, one can see that, although the spatial curvature can alleviate effectively the tension between different BAO data, it  leads to somewhat more serious tension between SNIa+BAO and CMB. After all, in a flat universe, SNIa+BAO3 and SNIa+BAO3+CMB give a very consistent constraint.  This result is in sharp contrast to that given in \cite{Cardenas} where only a CMB shift parameter is considered.  Since $R(z^{\ast})$, $l_{A}(z^{\ast})$ and $z^{\ast}$ give similar constraints on dark energy parameters compared with the full CMB power spectrum as is shown in~\cite{HongLi}, we think that our results may be more reliable than that reached in~\cite{Cardenas}.

\begin{acknowledgments}
We thank Lixin Xu for the help with the MCMC method.  This work was supported by
the National Natural Science Foundation of China under Grants Nos.
10935013, 11175093,  11222545  and 11075083, Zhejiang Provincial Natural Science
Foundation of China under Grants Nos. Z6100077 and R6110518, the
FANEDD under Grant No. 200922, the National Basic Research Program
of China under Grant No. 2010CB832803, the NCET under Grant No.
09-0144,  and K.C. Wong Magna Fund in Ningbo University.

\end{acknowledgments}


\begin{thebibliography}{99}
\bibitem{Perlmutter1999} S. Perlmutter,  G. Aldering, G. Goldhaber,  et al., Astrophys. J. 517, 565 (1999).
\bibitem{Riess1998} A. G. Riess,   A. V. Filippenko, P. Challis,  et al., Astron. J. 116, 1009 (1998).
\bibitem{Eisenstein}D. J. Eisenstein, et al., Astron. J. 633, 560 (2005).
\bibitem{Tegmark} M. Tegmark, et al., Phys. Rev. D 69, 103501 (2004).
\bibitem{Spergel} D. N. Spergel, et al., Astrophys. J. Suppl. Ser. 148, 175 (2003);
                            D. N. Spergel, et al., Astrophys. J. Suppl. Ser. 170, 377S (2007).
\bibitem{Hicken}      M. Hicken, W. M. Wood-Vasey, S. Blondin, P. Challis, S. Jha, P.L. Kelly, A. Rest,and R. P. Kirshner, Astrophys. J. 700, 1097 (2009).
\bibitem{Amanullah} R.  Amanullah, et al., Astrophys. J. 716, 712 (2010).
\bibitem{Percival2010}W. J. Percival,  et al., Mon. Not. R. Astron. Soc. 401, 2148 (2010);
                                  B.  A. Reid, et al., Mon. Not. Roy. Astron. Soc. 426, 2719 (2012).
\bibitem{Chevallier} M. Chevallier, D. Polarski, Int. J. Mod. Phys. D 10, 213 (2001);
                                E. V. Linder, Phys. Rev. Lett. 90, 091301 (2003).
\bibitem{Shafieloo} A. Shafieloo, V. Sahni, A.A. Starobinsky, Phys. Rev. D 80, 101301 (2009).
\bibitem{Li2011}   Z. Li, P. Wu and H. Yu, Phys. Lett. B 695, 1 (2011).
\bibitem{Gong} Y. Gong, B. Wang and R. Cai, J. Cosmol. Astropart. Phys 04, 019 (2010).
\bibitem{Li2010}Z. Li, P. Wu and H. Yu, J. Cosmol. Astropart. Phys 11,  031 (2010).
\bibitem{Guimaraes} A. C. C. Guimaraes and J. S. Lima,  Class. Quantum Grav. 28, 125026 (2011).
\bibitem{Cai}    R. Cai and Z. Tuo, Phys. Lett. B 706, 116 (2011);
                           P. Wu and H. Yu,   arXiv:1012.3032.
\bibitem{Hinshaw}G. Hinshaw et al., Astrophys. J. Suppl. Ser. 180, 225 (2009);
                             Y. Wang and P. Mukherjee, Phys. Rev. D 76, 103533 (2007).
\bibitem{Komatsu2011}E. Komatsu,  et al., Astrophys. J. Suppl. Ser. 180, 330 (2009).
\bibitem{Cardenas}V. H. Cardenas and M. Rivera,   Phys. Lett. B 710, 251 (2012).
\bibitem{Beutler} F.  Beutler, C. Blake, M. Colless, D. H. Jones, L. Staveley-Smith, L. Campbell, Q. Parker, W. Saunders,and F. Watson,  Mon. Not. R. Astron. Soc. 416,  3017 (2011).
\bibitem{Blake}   C. Blake,  et al.,  Mon. Not. R. Astron. Soc. 418, 1707 (2011).
\bibitem{Sanchez}A. G. Sanchez, et al., Mon. Not. R. Astron. Soc. 427,  3435 (2012).
\bibitem{HongLi} H. Li, J. Xia, G. Zhao, Z. Fan and X. Zhang, Astrophys. J. 683,  L1 (2008).
%\bibitem{Percivalb} W. J. Percival,  et al., Mon. Not. Roy. Astron. Soc.  401 (2010) 2148.
%\bibitem{Ariel2012}Ariel, G. et al. 2012, arXiv: 1203.6616v3.
\bibitem{Eisenstein1998}D. J. Eisenstein and W. Hu, Astrophys. J. 496,  605 (1998).
\bibitem{Hu1996}W. Hu and N. Sugiyama, Astrophys. J. 471,  542 (1996).
\bibitem{Lewis2002}A. Lewis and S. Bridle, Phys. Rev. D 66,  103511 (2002).

\end{thebibliography}
\end{document}